\documentclass[journal]{IEEEtran}

\usepackage{graphicx}
\usepackage{booktabs}
\usepackage{subfigure}
\usepackage{overpic}
\usepackage{amsmath}
\usepackage{flushend}
\usepackage{amssymb}
\usepackage{multirow}
\usepackage{makecell}
\usepackage{fontawesome}
\usepackage{color, soul}
\usepackage[ruled]{algorithm2e}
\usepackage{hyperref}
\hypersetup{
	colorlinks=true,
	linkcolor=blue,
	urlcolor=blue,
	citecolor=blue
}

\usepackage{colortbl}
\usepackage[dvipsnames]{xcolor}
\definecolor{bg}{HTML}{e0f1ff}

\begin{document}

\title{Semantic-Guided Fusion Network for Multi-source Remote Sensing Image Classification}

\author{
  Yuwei Zhao, 
  Chuanzheng Gong,
  Baogui Huan,
  Feng Gao, \emph{Member, IEEE},
  Junyu Dong, \emph{Member, IEEE},  \\
  and Qian Du, \emph{Fellow, IEEE}

  \thanks{
   This work was supported in part by National Key R \& D Program of China under Grant 2024YFF0908102, Key R \& D Program of Shandong Province under Grant 2025CXPT185, and in part by the Natural Science Foundation of Shandong Province under Grant ZR2024MF020. (\textit{Corresponding author: Feng Gao})

    Yuwei Zhao, Chuanzheng Gong, Baogui Huan, Feng Gao, and Junyu Dong are with the State Key Laboratory of Physical Oceanography, Ocean University of China, Qingdao 266100, China. (Email: gaofeng@ouc.edu.cn)

    Qian Du is with the Department of Electrical and Computer Engineering, Mississippi State University, Starkville, MS 39762 USA.
  }
}

\markboth{IEEE GEOSCIENCE AND REMOTE SENSING LETTERS}%
{Shell}

\maketitle

\label{sec:abstract}

\begin{abstract}
  Multi-source remote sensing image classification has attracted increasing attention due to the complementary spectral, structural, and geometric information. However, existing methods still suffer from two limitations: insufficient semantic contextual modeling and unreliable feature fusion caused by slight spatial misalignment. To address these issues, we propose a Semantic-Guided Fusion Network (SGFNet) for multi-source remote sensing image classification. Specifically, the \textit{Semantic Mixing Convolution Block (SMCB)} is designed to dynamically generate semantic-aware convolution kernels according to contextual relationships among feature representations. In addition, the \textit{Frequency Modulated Fusion Block (FMFB)} is introduced to perform cross-modal interaction in the frequency domain, which effectively alleviates the influence of slight spatial misalignment and improves complementary information fusion. Extensive experiments conducted on the Augsburg and Houston 2018 datasets demonstrate that the proposed SGFNet consistently outperforms several state-of-the-art methods. The codes are publicly available at \url{https://github.com/oucailab/SGFNet}.
\end{abstract}


\IEEEpeerreviewmaketitle

\section{Introduction}
\label{sec:intro}

\IEEEPARstart{W}{ith} the continuous growth of satellite missions and sensor technologies, massive volumes of high-resolution and multi-source remote sensing data are now available, providing unprecedented opportunities to monitor the Earth's surface in a timely and comprehensive manner \cite{dph26tip}. Hyperspectral images (HSIs) capture detailed spectral information across hundreds of contiguous wavelength bands. This rich spectral resolution makes HSI particularly valuable for land cover classification. However, HSI is susceptible to noise and atmospheric interference, which reduces data quality and reliability. To mitigate these challenges, light detection and ranging (LiDAR) and synthetic aperture radar (SAR) data are frequently integrated as complementary sources \cite{hxq26grsl}. LiDAR provides accurate, high-resolution 3D geometric and elevation information, while SAR offers valuable structural, scattering, and dielectric properties regardless of weather and lighting conditions. Consequently, the joint classification of HSI and LiDAR/SAR data has emerged as an effective paradigm to boost classification accuracy and robustness \cite{gf25msf}. Therefore, in this letter, we mainly focus on HSI and SAR/LiDAR joint classification. 

Deep learning-based methods have gained considerable research interest for multi-source remote sensing image joint classification, due to their powerful capability in learning hierarchical feature representations. By leveraging convolutional neural networks (CNNs) \cite{cnn_vit_backbones} \cite{gcz26tgrs}, Transformers \cite{zyk25grsl} \cite{cui2026mtmixer}, and more recently Mamba architectures \cite{mz26jstars}, these methods can effectively extract discriminative spatial-spectral features from HSI while simultaneously capturing complementary structural and geometric information from SAR and LiDAR data.

\begin{figure*}[!t]
  \centering
  \includegraphics[width=0.8\textwidth]{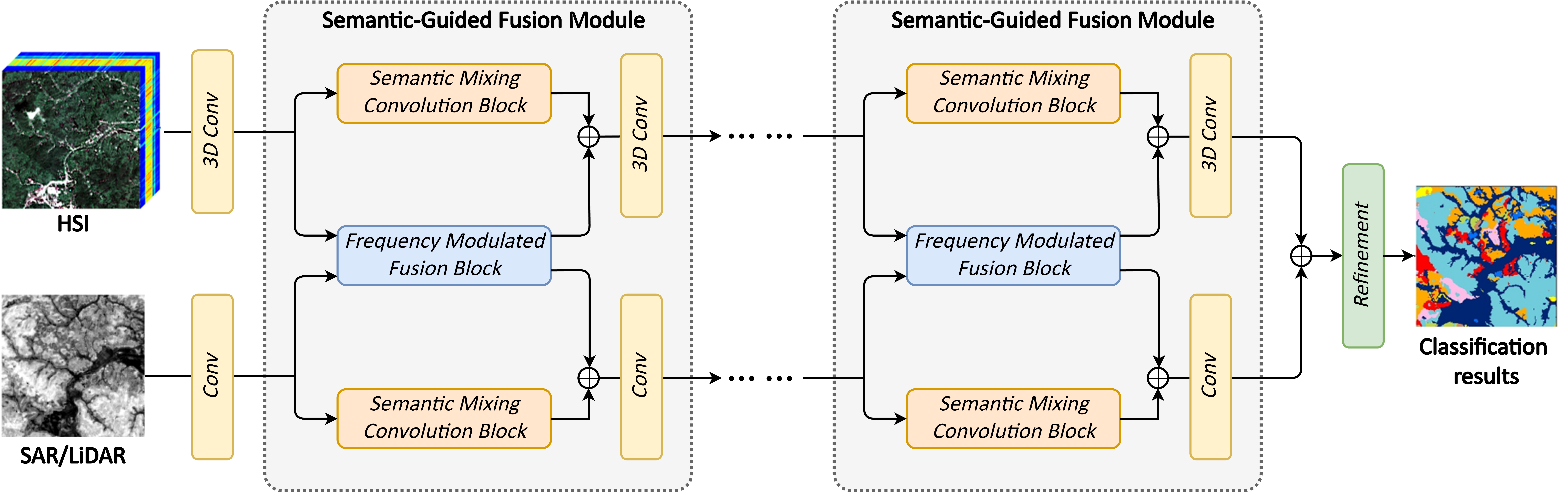}
  \caption{Overview of the Semantic-Guided Fusion Network (SGFNet). It adopts a dual-branch hierarchical architecture composed of multiple stacked Semantic-Guided Fusion Modules (SGFMs). Within each SGFM, two Semantic Mixing Convolution Blocks (SMCB) are introduced to model semantic-aware contextual representations. Frequency Modulated Fusion Block (FMFB) is inserted between the two branches.}
  \label{fig-frame}
\end{figure*}

Although existing methods have achieved remarkable classification performance, they still suffer from two limitations: \emph{\textbf{1) Lacking adaptive semantic and contextual representation capability.}} Traditional convolutions use spatially shared and fixed kernels for all spatial locations. Such fixed kernels can hardly adapt to varying semantic structures and content-aware feature distributions. \emph{\textbf{2) Slight cross-modal spatial misalignment weakens reliable feature fusion.}} Multi-source remote sensing data acquired by different sensors often suffer from slight spatial misalignment due to inconsistent imaging geometries and sensor characteristics. 

To effectively overcome these limitations, we propose the \textbf{S}emantic-\textbf{G}uided \textbf{F}usion \textbf{Net}work (\textbf{SGFNet}) for multi-source data joint classification. Specifically, to enhance the semantic and contextual feature representations for HSI and SAR/LiDAR data, we propose the \textit{Semantic Mixing Convolution Block (SMCB)}. It dynamically generates semantic-aware convolution kernels according to contextual relationships among feature representations. By enabling position-specific contextual modeling and selectively emphasizing semantically relevant information, the proposed module effectively enhances global semantic dependencies. In addition, we introduce the \textit{Frequency Modulated Fusion Block (FMFB)} to alleviate the slight spatial misalignment between multi-source data. It performs cross-modal feature interaction in the frequency domain rather than directly conducting spatial-domain fusion. Through adaptive frequency-aware modulation, our FMFB effectively enhances complementary information integration and improves the discriminative capability of multi-source feature representations.

Our main contributions are summarized as follows:

\begin{itemize}
  \item We propose SMCB to enhance semantic and contextual feature representations. Unlike conventional convolutions with spatially shared fixed kernels, our SMCB uses semantic-aware dynamic convolution to improve contextual dependency modeling.

  \item We introduce FMFB to alleviate the slight spatial misalignment. By performing adaptive cross-modal interaction in the frequency domain, our FMFB enhances complementary information integration. 
  
  \item Extensive experiments conducted on two benchmark datasets demonstrate that the proposed SGFNet consistently outperforms several state-of-the-art methods.
\end{itemize}

\section{Methodology}
\label{sec:method}

\subsection{Overall Framework of the Proposed SGFNet}

The overall framework of the proposed Semantic-Guided Fusion Network (SGFNet) is illustrated in Fig. \ref{fig-frame}. The framework adopts a dual-branch architecture, where the HSI and SAR/LiDAR data are first processed separately through convolutional layers to extract preliminary spectral–spatial and structural representations. To achieve effective cross-modal interaction, the network introduces multiple stacked Semantic-Guided Fusion Modules (SGFM). Within each module, a Semantic Mixing Convolution Block (SMCB) is employed to capture high-level semantic contextual information and enhance discriminative feature learning. Meanwhile, a Frequency Modulated Fusion Block (FMFB) is designed to model complementary frequency-domain characteristics between heterogeneous modalities, enabling the network to preserve fine-grained structural details and suppress redundant or noisy information. By progressively stacking several SGFMs, the framework performs hierarchical semantic refinement and deep multi-modal feature fusion. Finally, the fused representations are integrated through two convolutional layers for refinement to generate the final classification results.

\begin{figure}[!t]
  \centering
  \includegraphics[width=3in]{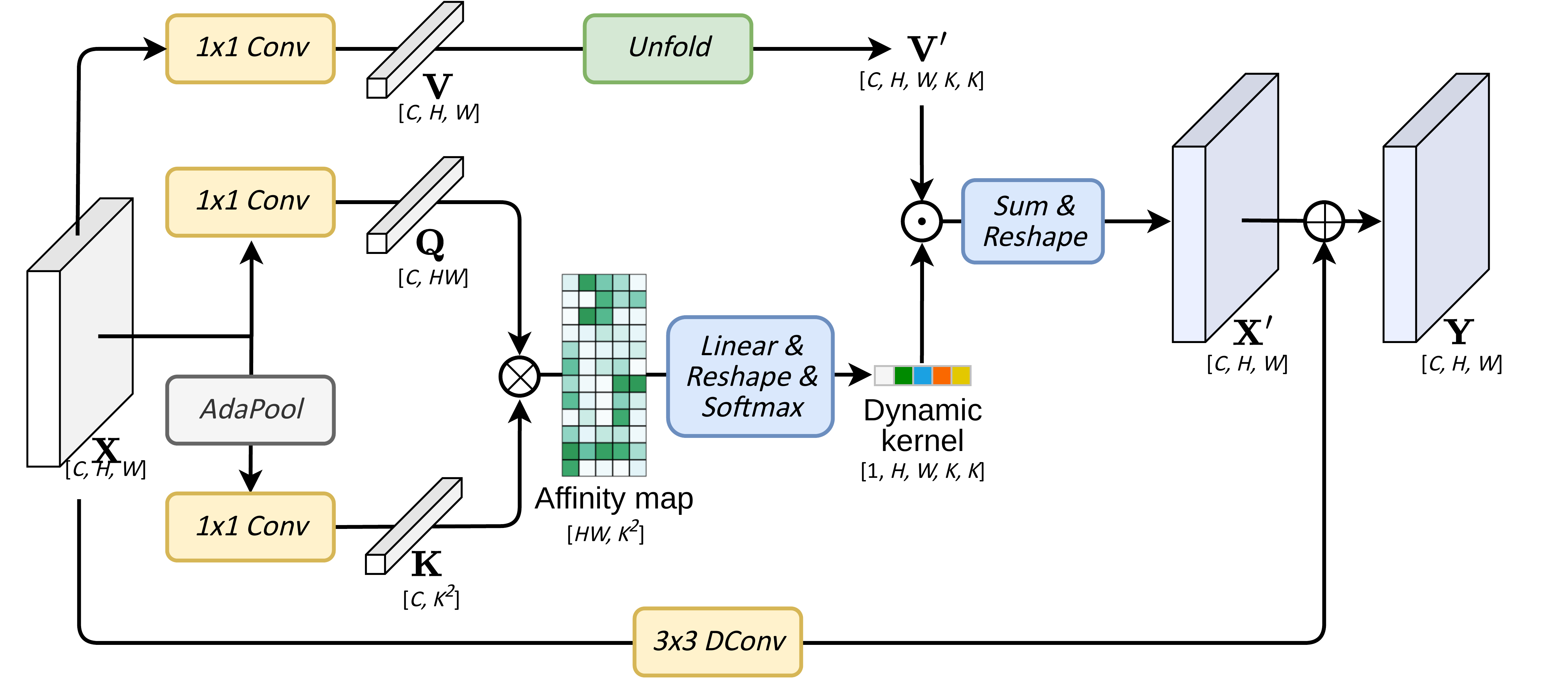}
  \caption{Illustration of the Semantic Mixing Convolution Block (SMCB).}
  \label{fig-smcb}
\end{figure}

\subsection{Semantic Mixing Convolution Block (SMCB)}

Standard convolutions use spatially shared fixed kernels, limiting their ability to adapt to varying semantic structures and contextual relationships. To address this issue, we propose the Semantic Mixing Convolution Block (SMCB), which integrates convolution with semantic-aware dynamic filtering. Details of the SMCB are shown in Fig. \ref{fig-smcb}. Given an input feature $\mathbf{X}\in\mathbb{R}^{C\times H\times W}$, it is projected by a $1\times1$ convolution to obtain the value and query features as: 
\begin{equation}
\mathbf{V}=f_v(\mathbf{X}), ~~~ \mathbf{Q}=f_q(\mathbf{X}),
\end{equation}
where $f_v$ and $f_q$ denote the convolution layer. Meanwhile, $\mathbf{X}$ is passed through an adaptive pooling layer to extract compact contextual information as:
\begin{equation}
  \mathbf{K} = f_k(\mathrm{AdaPool}(\mathbf{X})),
\end{equation}
where $f_k$ denotes the convolution layer. We obtain $\mathbf{Q}\in\mathbb{R}^{C\times HW}$ and $\mathbf{K}\in\mathbb{R}^{C\times K^2}$ via reshaping. Here $K$ denotes the dynamic kernel size. The query and key features are multiplied to obtain an affinity map:
\begin{equation}
\mathbf{A}=\mathbf{Q}^{T}\mathbf{K}.
\end{equation}

This affinity map measures the relationship between each spatial position and the $K^2$ sampling locations in a local window. It is then processed by a linear layer, reshaped, and normalized by $\mathrm{Softmax}$ to generate the dynamic convolution kernel:
\begin{equation}
\mathbf{D}=\mathrm{Softmax}(\mathrm{Reshape}(\mathrm{Linear}(\mathbf{A}))).
\end{equation}

In parallel, the value feature $\mathbf{V}$ is unfolded into local patches:
\begin{equation}
\mathbf{V}'=\operatorname{Unfold}(\mathbf{V}).
\end{equation}

The dynamic kernel $\mathbf{D}$ is then applied to the unfolded value feature. In each channel of $\mathbf{V'}$, for every spatial position $(h,w)$, the output feature is computed by weighted summation over its local $K\times K$ neighborhood.

Besides the dynamic context branch, the framework also includes a local feature enhancement branch. The original input $\mathbf{X}$ is processed by a $3\times3$ depth-wise convolution to compute $\mathbf{X}_d$. Finally, the dynamically aggregated feature $\mathbf{X}'$ is combined with the local depth-wise convolution feature through residual addition to produce the final output.

\subsection{Frequency Modulated Fusion Block}

\begin{figure}[!t]
  \centering
  \includegraphics[width=2.5in]{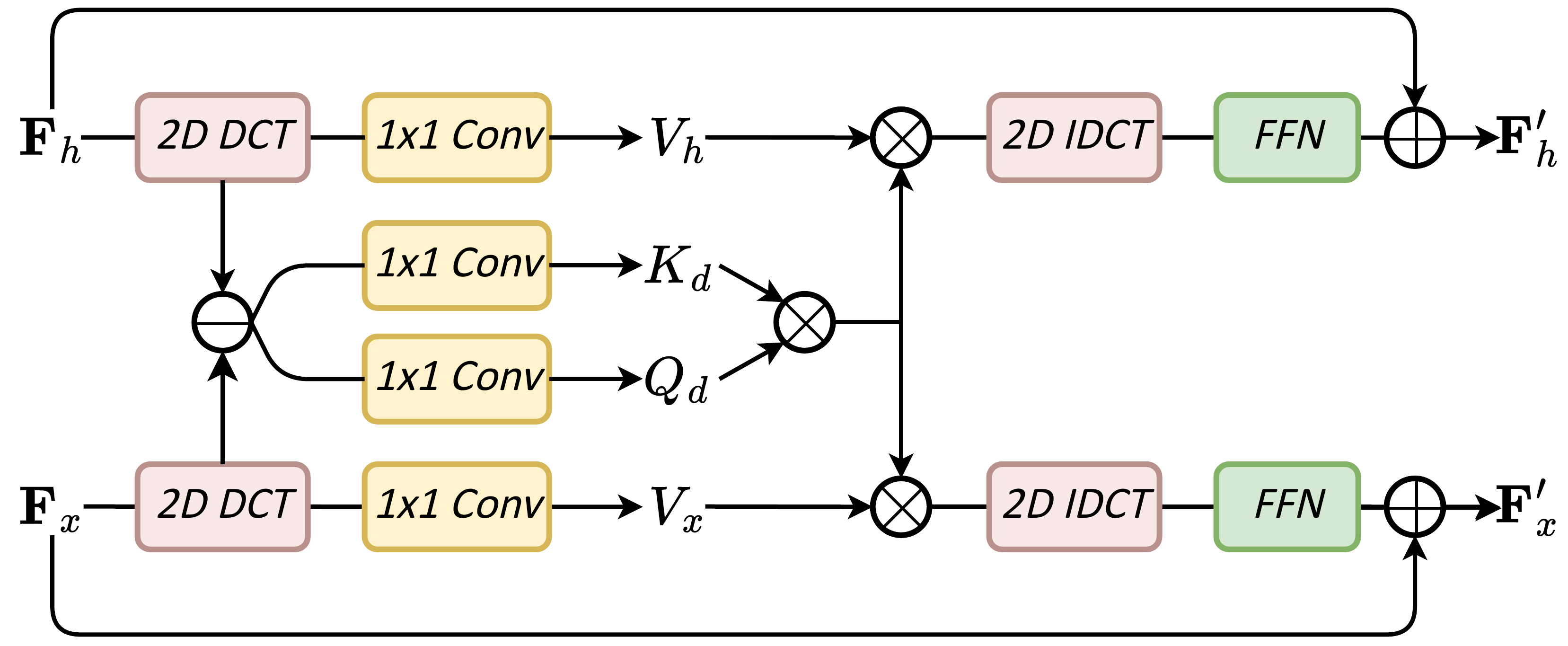}
  \caption{Illustration of the Frequency Modulated Fusion Block (FMFB).}
  \label{fig-fmfb}
\end{figure}

Multi-source data are usually collected by different sensors with varying imaging geometries and acquisition times, which often results in slight spatial misalignment between modalities. Local object regions may exhibit small positional deviations and structural inconsistencies. Such slight misalignment may weaken the effectiveness of cross-modal feature fusion. To this end, we propose the Frequency Modulated Fusion Block (FMFB), which performs adaptive cross-modal fusion in the frequency domain to enhance feature consistency.

Details of FMFB are illustrated in Fig. \ref{fig-fmfb}. It takes the feature representations $\mathbf{F}_h$ from the HSI branch and $\mathbf{F}_x$ from the SAR/LiDAR branch as inputs. It first applies 2D Discrete Cosine Transform (DCT) to project both features into the frequency domain:
\begin{equation}
\hat{\mathbf{F}}_h=\mathrm{DCT}(\mathbf{F}_h),
\quad
\hat{\mathbf{F}}_x=\mathrm{DCT}(\mathbf{F}_x).
\end{equation}

We then compute the element-wise difference in the frequency domain:
\begin{equation}
\hat{\mathbf{F}}_d=\hat{\mathbf{F}}_h-\hat{\mathbf{F}}_x.  
\end{equation}

Based on this joint representation, two $1\times1$ convolution layers generate the query and key features as $\mathbf{Q}_d=f_q(\hat{\mathbf{F}}_d)$, $\mathbf{K}_d=f_k(\hat{\mathbf{F}}_d)$.  Meanwhile, both DCT-transformed inputs are separately projected into value features:
\begin{equation}
\mathbf{V}_h=f_v^h(\hat{\mathbf{F}}_h),
\quad
\mathbf{V}_x=f_v^x(\hat{\mathbf{F}}_x).
\end{equation}

Next, attention is computed as follows:
\begin{equation}
 \mathbf{F}^{attn}_{h} = \mathrm{Softmax}\left(
 \frac{\mathbf{Q}_d \mathbf{K}_d^{\top}}{\sqrt{d_k}}
 \right)\mathbf{V}_h,
\end{equation}
\begin{equation}
 \mathbf{F}^{attn}_{x} = \mathrm{Softmax}\left(
 \frac{\mathbf{Q}_d \mathbf{K}_d^{\top}}{\sqrt{d_k}}
 \right) \mathbf{V}_x,
\end{equation}
where $d_k$ denotes the channel dimension of $\mathbf{Q}_d$ and $\mathbf{K}_d$, used as a scaling factor to stabilize gradients during training.

After that, the modulated frequency-domain features are transformed back into the spatial domain through 2D inverse DCT:
\begin{equation}
\hat{\mathbf{F}}'_h=\operatorname{IDCT}(\mathbf{F}^{attn}_h),
\quad
\hat{\mathbf{F}}'_x=\operatorname{IDCT}(\mathbf{F}^{attn}_x).
\end{equation}

The reconstructed feature is further refined by the Feed-Forward Network (FFN) and residual connections:
\begin{equation}
\mathbf{F}'_h=\operatorname{FFN}(\hat{\mathbf{F}}'_h) + \mathbf{F}_h,
\quad
\mathbf{F}'_x=\operatorname{FFN}(\hat{\mathbf{F}}'_x) + \mathbf{F}_x.
\end{equation}

\begin{equation}
\mathbf{F}'_h=\operatorname{FFN}(\hat{\mathbf{F}}'_h) + \mathbf{F}_h,
\quad
\mathbf{F}'_x=\operatorname{FFN}(\hat{\mathbf{F}}'_x) + \mathbf{F}_x.
\end{equation}

Frequency-domain interaction is motivated by the shift theorem: small spatial translations mainly affect phase while largely preserving the frequency-energy distribution. Therefore, FMFB is less sensitive to slight cross-modal misalignment than pixel-wise spatial fusion.

\section{Experimental Results and Analysis}
\label{sec:results_analysis}

\subsection{Datasets and Experimental Settings}

\begin{figure*}[!t]
  \centering
  \includegraphics[width=0.8\textwidth]{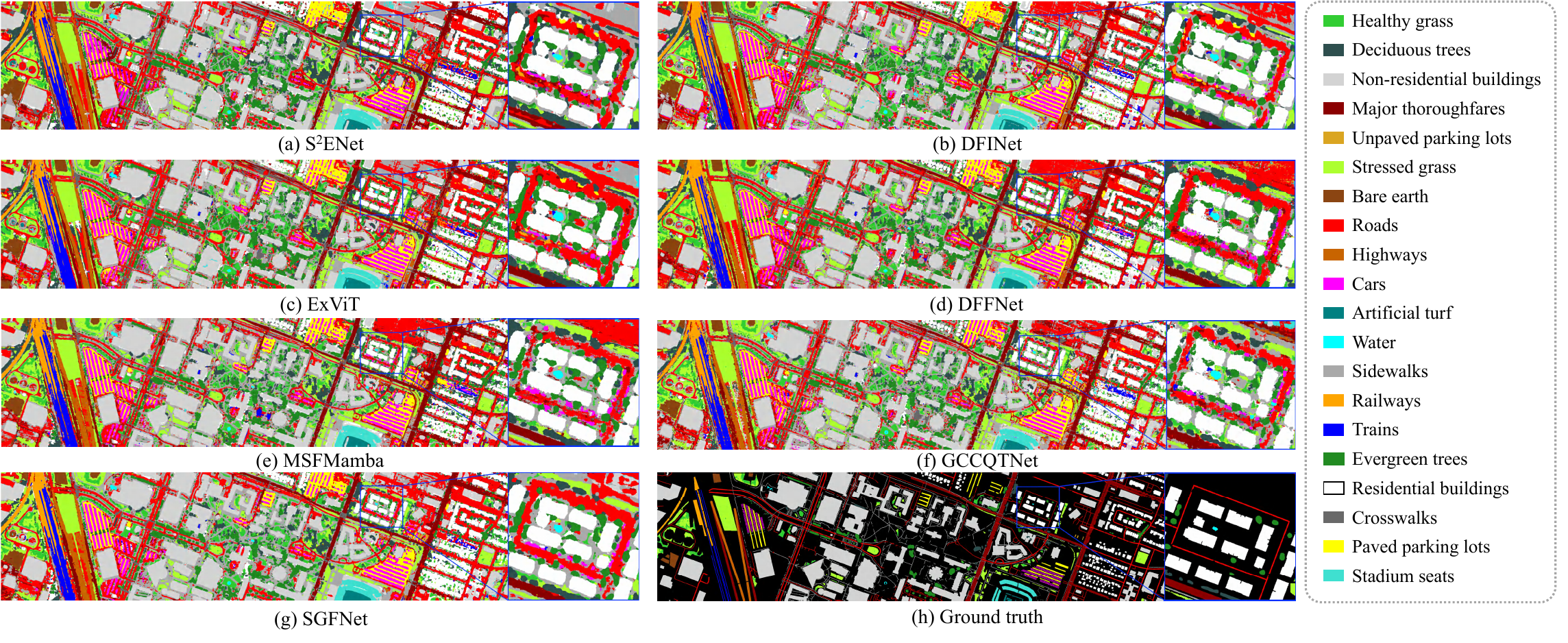}
  \caption{Classification results of different methods on the Houston2018 dataset.}
  \label{fig_results}
\end{figure*}

\begin{table}
  \centering
  \caption{Experimental results on the Augsburg dataset. The \textbf{bold} and \underline{underline} denote the best and second results.}
  \setlength{\tabcolsep}{3pt} 
  \resizebox{0.90\linewidth}{!}{
  \begin{tabular}{c|ccccccc}
  \hline\toprule
  Class            & S$^2$ENet & DFINet & ExViT & DFFNet & MSFM & GCCQTNet & \cellcolor{bg}SGFNet \\
  \midrule
  Forest           & \textbf{98.10} & \underline{97.38} & 90.04 & 96.17 & 97.17 & 95.26 & \cellcolor{bg}97.18 \\
  Residential area & \textbf{99.08} & \underline{98.37} & 95.44 & 96.64 & 98.15 & 96.80 & \cellcolor{bg}98.28 \\
  Industrial area  & 12.19 & 61.31 & 34.58 & 37.70 & 50.26 & \textbf{68.20} & \cellcolor{bg}\underline{61.41} \\
  Low plants       & 91.78 & 92.63 & 90.68 & 94.21 & \textbf{95.52} & 91.35 & \cellcolor{bg}\underline{95.48} \\
  Allotment        & 45.12 & 49.33 & 51.82 & \underline{53.54} & 53.35 & 41.68 & \cellcolor{bg}\textbf{65.01} \\
  Commercial area  & 1.22  & 3.54  & \textbf{28.63} & 11.36 & 2.63 & \underline{13.97} & \cellcolor{bg}8.55 \\
  Water            & 24.09 & 26.61 & 17.65 & 27.15 & \underline{49.97} & 46.80 & \cellcolor{bg}\textbf{56.62} \\
  \midrule
  OA   & 88.22 & 90.66 & 86.65 & 89.37 & \underline{91.38} & 90.19 & \cellcolor{bg}\textbf{92.38} \\
  AA               & 53.08 & 61.31 & 58.41 & 59.54 & 63.31 & \underline{64.87} & \cellcolor{bg}\textbf{68.93} \\
  Kappa            & 86.47 & 86.47 & 80.79 & 84.51 & \underline{87.45} & 85.85 & \cellcolor{bg}\textbf{89.05} \\
  \bottomrule\hline
  \end{tabular}}
  \label{table_augsburg}
\end{table}

The performance of the proposed SGFNet is evaluated on two benchmark multi-source remote sensing datasets for land-cover classification. The first dataset is the Augsburg Dataset \cite{aug23data}, which consists of HSI and SAR data. It contains ground-truth annotations for seven land-cover classes. The second dataset is the Houston 2018 Dataset, which comprises HSI and LiDAR data. As it is obtained from the IEEE GRSS Data Fusion Contest 2018 \cite{xyh18grss}, this dataset provides a representative multi-sensor optical geospatial benchmark for challenging urban land-cover and land-use classification tasks over the University of Houston campus and its surrounding areas. 

Our SGFNet was conducted on NVIDIA GeForce RTX 5080 GPU. The training phase spanned over 100 epochs. The AdamW optimizer was used with a learning rate of 0.001. The batch size was set to 32. To evaluate the performance of the proposed SGFNet, we compare it against other state-of-the-art methods, including S$^2$ENet \cite{fs22grsl}, DFINet \cite{gyh22tgrs}, ExViT \cite{exvit}, DFFNet \cite{dffnet}, and MSFM \cite{gao2025msfmamba}. To quantitatively evaluate the experimental results, we employ three standard metrics: Overall Accuracy (OA), Average Accuracy (AA), and Kappa coefficient. 

\subsection{Parameter Analysis}

\begin{figure}[htb]
\centering
\includegraphics[width=3in]{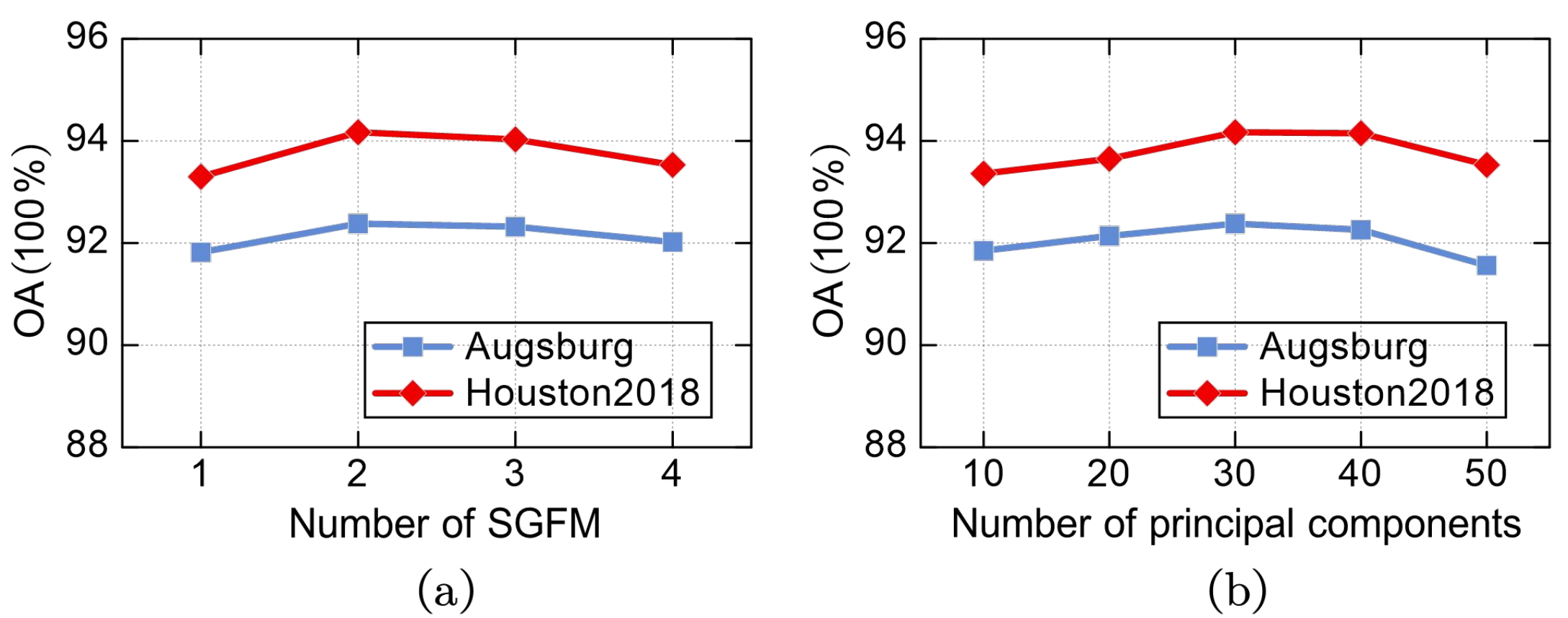}
\caption{Parameter analysis. (a) The relationship between OA and the number of SGFM. (b) The relationship between OA and the number of principal components after PCA for HSI.}
  \label{fig_para}
\end{figure}

\textbf{Number of SGFM.} We test different numbers of SGFM from 1 to 4. The experimental results are shown in Fig. \ref{fig_para}(a). When the number of SGFM is set to 2, our SGFNet achieves the best performance on both datasets. Therefore, in the following experiments, we use two SGFMs in the proposed SGFNet.

\textbf{Number of Principal Components for HSI.} We use Principal Component Analysis (PCA) for HSI preprocessing. We tested the number of principal components for HSI, and the experimental results are shown in Fig. \ref{fig_para}(b). It can be observed that when the number of principal components is set to 30, our SGFNet achieves the best classification performance. Retaining too few components discards crucial discriminative spectral features, leading to suboptimal accuracy. Thus, the number of principal components is empirically set to 30.

\textbf{Sensitivity of the Dynamic Kernel Size in SMCB.} We evaluate SGFNet with $K \in \{5, 7, 9, 11\}$ on both datasets, and the results are shown in Table \ref{table_kernel}. The OA on both datasets peaks at $K=7$. A smaller $K$ provides insufficient semantic sampling locations for effective contextual modeling, while a larger $K$ introduces potentially redundant or noisy contextual information. Hence, we set $K=7$ in our experiments, as it achieves the best trade-off between classification accuracy and computational overhead.

\begin{table}[h]
  \centering
  \caption{OA (\%) of SGFNet under different dynamic kernel sizes $K$ in SMCB.}
  \resizebox{0.65\columnwidth}{!}{
  \begin{tabular}{lcccc}
  \toprule
  $K$ & 5 & 7 & 9 & 11 \\
  \midrule
  Augsburg    & 91.64 & \textbf{92.58} & 92.50 & 92.02 \\
  Houston2018 & 93.26 & \textbf{94.10} & 94.03 & 93.47 \\
  \bottomrule
  \end{tabular}}
  \label{table_kernel}
\end{table}

\subsection{Experimental Results and Analysis}

\textbf{Results on the Augsburg Dataset.} Table \ref{table_augsburg} shows the quantitative evaluations on the Augsburg dataset. The OA value of the proposed SGFNet achieves 92.38\%, outperforming all the other methods. From the class-wise results, SGFNet achieves the best performance in Allotment, and remains highly competitive in the Industrial area. Both categories usually exhibit complex spatial structures and significant spectral ambiguity, indicating that SGFNet can effectively capture discriminative semantic and contextual information.

\begin{table}
  \centering
  \caption{Experimental results on the Houston2018 dataset. The \textbf{bold} and \underline{underline} denote the best and second best results. }
  \setlength{\tabcolsep}{2pt} 
  \resizebox{0.9\linewidth}{!}{
  \begin{tabular}{c|ccccccc}
  \hline\toprule
  Class & S$^2$ENet & DFINet & ExViT & DFFNet & MSFM & GCCQTNet & \cellcolor{bg} SGFNet \\
  \midrule
  Healthy grass                 & \textbf{99.15} & \underline{94.29} & 93.34 & 92.63 & 90.38 & 91.71 & \cellcolor{bg} 92.07 \\
  Stressed grass               & 84.35 & 92.71 & 93.44 & 92.44 & \underline{94.04} & \textbf{94.22} & \cellcolor{bg} 93.18 \\
  Artificial turf              & \textbf{100.0}   & \textbf{100.0}   & \textbf{100.0} & \textbf{100.0} & \textbf{100.0} & \textbf{100.0} & \cellcolor{bg} \textbf{100.0} \\
  Evergreen trees              & 98.13 & 98.92 & 95.87 & \underline{99.38} & 99.34 & 98.58 & \cellcolor{bg} \textbf{99.76} \\
  Deciduous trees              & 86.39 & 97.77 & 96.85 & \underline{99.16} & 99.11 & 98.71 & \cellcolor{bg} \textbf{99.94} \\
  Bare earth                   & 99.33 & \textbf{100.0}   & \textbf{100.0} & \underline{99.96} & \textbf{100.0} & \underline{99.96} & \cellcolor{bg} \textbf{100.0} \\
  Water & \textbf{100.0}   & \textbf{100.0}   & \textbf{100.0} & \textbf{100.0} & \textbf{100.0} & \textbf{100.0} & \cellcolor{bg} \textbf{100.0} \\
  Residential buildings        & 97.17 & 97.36 & 97.39 & \underline{98.00} & 96.27 & 96.41 & \cellcolor{bg} \textbf{98.78} \\
  Non-residential buildings    & 93.93 & 93.29 & 93.00 & 93.73 & \underline{95.58} & 94.16 & \cellcolor{bg} \textbf{96.24} \\
  Roads  & 70.17 & 75.73 & 72.18 & \underline{80.31} & 75.90 & 77.79 & \cellcolor{bg} \textbf{85.33} \\
  Sidewalks                    & 72.11 & \underline{83.67} & 69.69 & 72.59 & 81.43 & 82.55 & \cellcolor{bg} \textbf{84.64} \\
  Crosswalks                   & 82.47 & 94.23 & 88.12 & 91.84 & 96.22 & \underline{97.11} & \cellcolor{bg} \textbf{98.14} \\
  Major thoroughfares          & \textbf{89.44} & 81.20 & 86.12 & 84.54 & 86.89 & 88.03 & \cellcolor{bg} \underline{88.85} \\
  Highways                     & 98.53 & 98.91 & \textbf{99.50} & 98.82 & 98.63 & 98.43 & \cellcolor{bg} \underline{99.35} \\
  Railways                     & 99.21 & \textbf{99.94} & 99.87 & \underline{99.89} & 99.69 & 99.77 & \cellcolor{bg} 99.82 \\
  Paved parking lots           & 95.62 & 98.37 & 97.02 & 97.78 & 98.94 & \textbf{99.35} & \cellcolor{bg}\underline{99.34} \\
  Unpaved parking lots         & \textbf{100.0}   & \textbf{100.0}   & \textbf{100.0} & \textbf{100.0} & \textbf{100.0} & \textbf{100.0} & \cellcolor{bg}\textbf{100.0} \\
  Cars                         & 96.77 & \textbf{99.09} & 98.13 & \underline{98.89} & 97.81 & 97.44 & \cellcolor{bg}98.67 \\
  Trains                       & \textbf{100.0}   & 99.46 & \underline{99.98} & 99.91 & 99.97 & 99.89 & \cellcolor{bg}\textbf{100.0} \\
  Stadium seats                & \textbf{100.0}   & 99.98 & \underline{99.99} & \textbf{100.0} & \textbf{100.0} & 99.84 & \cellcolor{bg}\textbf{100.0} \\
  \midrule
  OA                           & 90.05 & 91.02 & 89.98 & 91.21 & \underline{92.38} & 92.13 & \cellcolor{bg}\textbf{94.17} \\
  AA                           & 93.14 & 95.24 & 94.02 & 94.99 & 95.51 & \underline{95.70} & \cellcolor{bg}\textbf{96.71} \\
  Kappa & 87.20 & 88.46 & 87.14 & 88.68 & \underline{90.16} & 89.86 & \cellcolor{bg}\textbf{92.46} \\
  \bottomrule\hline
  \end{tabular}}
  \label{table_houston2018}
\end{table}

\textbf{Results on the Houston2018 Dataset.} The classification maps on the Houston2018 dataset are shown in Fig. \ref{fig_results}, and the corresponding quantitative evaluations are illustrated in Table \ref{table_houston2018}. It can be observed that SGFNet produces classification maps that are more consistent with the ground truth, demonstrating superior land-cover discrimination capability. It shows that our SGFNet obtains the highest OA and Kappa, significantly outperforming other state-of-the-art methods. SGFNet achieves the best performance in several challenging categories, including Evergreen trees, Deciduous trees, Buildings, Roads, Sidewalks, Crosswalks, and Paved parking lots. These categories usually contain complex spatial distributions and strong spectral similarity with surrounding objects. The superior performance of SGFNet indicates that the proposed framework can effectively capture discriminative semantic information from multi-source data.

Although SGFNet is not specifically designed for class-imbalance learning, it partially alleviates the negative effects of imbalanced samples. SGFNet achieves notable improvements on minority classes (Roads, Sidewalks, Crosswalks, and Residential buildings), together with the highest AA of 96.71\%. It indicates that its semantic feature extraction and cross-modal fusion enhance minority-class discrimination. However, without balanced sampling, loss reweighting, or data augmentation, SGFNet mitigates rather than fundamentally solves the class-imbalance problem.

\begin{table}[!t]
\centering
\caption{Ablation study of the proposed SGFNet.}
\label{table_ablation}
\resizebox{0.65\linewidth}{!}{
\begin{tabular}{c|cc} 
\toprule
Model Variant & Augsburg & Houston2018 \\ 
\midrule
Baseline   & 88.54 & 89.65 \\
w/o SMCB   & 89.88 & 91.01 \\
w/o FMFB   & 91.29 & 92.37 \\ 
\rowcolor{bg}SGFNet (Full model) & \textbf{92.38} & \textbf{94.17} \\
\bottomrule
\end{tabular}}
\end{table}

\subsection{Ablation Study}

Table \ref{table_ablation} presents the ablation study results of the proposed SGFNet. The experiments are designed to evaluate the effectiveness of SMCB and FMFB. Both modules consistently improve the classification performance, it demonstrates that SMCB can effectively enhance contextual representation capability, and FMFB effectively alleviates the slight spatial misalignment between multi-source data. The complete SGFNet achieves significant improvements of 3.84\% and 4.52\%, respectively. These results demonstrate that both proposed modules are complementary and jointly contribute to more robust classification. 

\subsection{Robustness to Spatial Misalignment}

We conduct experiments on the Houston2018 dataset by applying spatial shifts of $\{1, 2, 3, 5\}$ pixels along both the $x$- and $y$-directions to the LiDAR branch input, while keeping the HSI branch fixed. As shown in Table \ref{table_alignment}, the OA of DFFNet and MSFMamba degrades by 2.60\% and 2.03\%, respectively, whereas SGFNet exhibits only 1.40\% drop. It confirms that the proposed FMFB alleviates cross-modal spatial misalignment through frequency-domain modulation.

\begin{table}[h]
  \centering
  \caption{OA (\%) under spatial shifts on the Houston2018 dataset.}
  \resizebox{0.65\columnwidth}{!}{
  \begin{tabular}{lccccc}
  \toprule
  Shift (pixels) & 0 & 1 & 2 & 3 & 5 \\
  \midrule
  DFFNet & 91.21 & 90.98 & 90.09 & 89.36 & 88.61 \\
  MSFMamba & 92.38 & 92.14 & 92.07 & 91.81 & 90.35 \\
  \rowcolor{bg} \textbf{SGFNet (Ours)} & \textbf{94.17} & \textbf{94.03} & \textbf{93.96} & \textbf{93.64} & \textbf{92.77} \\
  \bottomrule
  \end{tabular}}
  \label{table_alignment}
\end{table}

\section{Conclusions}
\label{sec:conclusion}

In this letter, we propose SGFNet for multi-source remote sensing image classification. To address the limitation of traditional convolutions in modeling semantic contextual relationships, we propose Semantic Mixing Convolution Block (SMCB) to dynamically generate semantic-aware convolution kernels according to contextual affinities among feature representations. In addition, we introduce the Frequency Modulated Fusion Block (FMFB) to perform adaptive cross-modal interaction in the frequency domain. Extensive experiments on two benchmark datasets demonstrate that SGFNet consistently outperforms state-of-the-art methods.

\bibliographystyle{IEEEtran}
\bibliography{references}

@article{dph26tip,
  author={Duan, Puhong and Jin, Shiyu and Lu, Xiaotian and Liang, Lianhui and Kang, Xudong and Plaza, Antonio},
  journal={IEEE Transactions on Image Processing}, 
  title={Domain-Adaptive Mamba for Cross-Scene Hyperspectral Image Classification}, 
  year={2026},
  volume={35},
  pages={1206-1217}}

@article{hxq26grsl,
  author={He, Xueqin and Xu, Kaijian and Wang, Shuzhou and Zhao, Ping and Zhang, Yuqi and Bi, Juanjuan},
  journal={IEEE Geoscience and Remote Sensing Letters}, 
  title={MUFNet: A Mamba-Based Uncertainty-Aware Fusion Framework for Fine-Grained Satellite Hyperspectral-LiDAR Tree Species Mapping}, 
  year={2026},
  volume={23},
  pages={1-5}}

@ARTICLE{mz26jstars,
  author={Meng, Zhe and Yang, Zixin and Zhao, Feng},
  journal={IEEE Journal of Selected Topics in Applied Earth Observations and Remote Sensing}, 
  title={Joint Classification of Hyperspectral and LiDAR Data Based on Dynamic Sparse-Dense Transformer and Mamba}, 
  year={2026},
  volume={19},
  pages={11445-11461}}

@article{gcz26tgrs,
  author={Gong, Chuanzheng and Gao, Feng and Lin, Junyan and Dong, Junyu and Du, Qian},
  journal={IEEE Transactions on Geoscience and Remote Sensing}, 
  title={Representative Spectral Correlation Network for Multisource Remote Sensing Image Classification}, 
  year={2026},
  volume={64},
  pages={1-14}}

@article{zyk25grsl,
  author={Zhao, Yikang and Gao, Feng and Jin, Xuepeng and Dong, Junyu and Du, Qian},
  journal={IEEE Geoscience and Remote Sensing Letters}, 
  title={Dynamic Frequency Feature Fusion Network for Multisource Remote Sensing Data Classification}, 
  year={2025},
  volume={22},
  pages={1-5}}

@article{gf25msf,
  author={Gao, Feng and Jin, Xuepeng and Zhou, Xiaowei and Dong, Junyu and Du, Qian},
  journal={IEEE Transactions on Geoscience and Remote Sensing}, 
  title={{MSFMamba}: Multiscale Feature Fusion State Space Model for Multi-source Remote Sensing Image Classification}, 
  year={2025},
  volume={63},
  pages={1-16}}

@article{cui2026mtmixer,
  title={{MTMixer}: a hybrid {Mamba-Transformer} architecture for multimodal remote sensing image classification},
  author={Cui, Xiandai and Zhang, Li},
  journal={International Journal of Remote Sensing},
  volume={47},
  number={2},
  pages={770--799},
  year={2026}}

@article{cnn_vit_backbones,
  author={Li, Ke and Wang, Di and Wang, Xu and Liu, Gang and Wu, Zili and Wang, Quan},
  journal={IEEE Transactions on Geoscience and Remote Sensing}, 
  title={Mixing Self-Attention and Convolution: A Unified Framework for Multisource Remote Sensing Data Classification}, 
  year={2023},
  volume={61},
  pages={1-16}}

@article{fs22grsl,
  author={Fang, Sheng and Li, Kaiyu and Li, Zhe},
  journal={IEEE Geoscience and Remote Sensing Letters}, 
  title={{S2ENet}: Spatial–Spectral Cross-Modal Enhancement Network for Classification of Hyperspectral and {LiDAR} Data}, 
  year={2022},
  volume={19},
  pages={1-5}}

@article{gyh22tgrs,
  author={Gao, Yunhao and Li, Wei and Zhang, Mengmeng and Wang, Jianbu and Sun, Weiwei and Tao, Ran and Du, Qian},
  journal={IEEE Transactions on Geoscience and Remote Sensing}, 
  title={Hyperspectral and Multispectral Classification for Coastal Wetland Using Depthwise Feature Interaction Network}, 
  year={2022},
  volume={60},
  pages={1-15}}

@article{dffnet,
  author={Zhao, Yikang and Gao, Feng and Jin, Xuepeng and Dong, Junyu and Du, Qian},
  journal={IEEE Geoscience and Remote Sensing Letters}, 
  title={Dynamic Frequency Feature Fusion Network for Multisource Remote Sensing Data Classification}, 
  year={2025},
  volume={22},
  number={},
  pages={1-5},
  doi={10.1109/LGRS.2025.3586958}}

@article{gao2025msfmamba,
  title={MSFMamba: Multi-scale feature fusion state space model for multi-source remote sensing image classification},
  author={Gao, Feng and Jin, Xuepeng and Zhou, Xiaowei and Dong, Junyu and Du, Qian},
  journal={IEEE Transactions on Geoscience and Remote Sensing},
  year={2025},
  publisher={IEEE}
}

@article{exvit,
  author={Yao, Jing and Zhang, Bing and Li, Chenyu and Hong, Danfeng and Chanussot, Jocelyn},
  journal={IEEE Transactions on Geoscience and Remote Sensing}, 
  title={Extended Vision Transformer (ExViT) for Land Use and Land Cover Classification: A Multimodal Deep Learning Framework}, 
  year={2023},
  volume={61},
  pages={1-15}}

@article{aug23data,
AUTHOR = {Hu, J. and Liu, R. and Hong, D. and Camero, A. and Yao, J. and Schneider, M. and Kurz, F. and Segl, K. and Zhu, X. X.},
TITLE = {MDAS: a new multimodal benchmark dataset for remote sensing},
JOURNAL = {Earth System Science Data},
VOLUME = {15},
YEAR = {2023},
NUMBER = {1},
PAGES = {113--131}}

@article{xyh18grss,
  author={Xu, Yonghao and Du, Bo and Zhang, Liangpei and Cerra, Daniele and Pato, Miguel and Carmona, Emiliano and Prasad, Saurabh and Yokoya, Naoto and Hänsch, Ronny and Le Saux, Bertrand},
  journal={IEEE Journal of Selected Topics in Applied Earth Observations and Remote Sensing}, 
  title={Advanced Multi-Sensor Optical Remote Sensing for Urban Land Use and Land Cover Classification: Outcome of the 2018 IEEE GRSS Data Fusion Contest}, 
  year={2019},
  volume={12},
  number={6},
  pages={1709-1724}}

\end{document}